\documentclass[aps,pre,twocolumn,nofootinbib,floatfix]{revtex4}
\usepackage{amsmath}
\usepackage{epsfig}

\input{tcilatex}

\begin{document}

\title{Bound states of L-shaped or T-shaped quantum wires in inhomogeneous
magnetic fields}
\author{Yuh-Kae Lin, Yueh-Nan Chen and Der-San Chuu\thanks{%
Corresponding author email address: dschuu@cc.nctu.edu.tw;
Fax:886-3-5725230; Tel:886-3-5712121-56105.}}
\affiliation{Department of Electrophysics, National Chiao Tung University, Hsinchu, Taiwan}
\date{\today }

\begin{abstract}
The bound state energies of L-shaped or T-shaped quantum wires in
inhomogeous magnetic fields are found to depend strongly on the asymmetric
parameter $\alpha =W_{2}/W_{1}$, i.e. the ratio of the arm widths. Two
effects of magnetic field on bound state energies of the electron are
obtained. One is the depletion effect which purges the electron out of the
OQD system. The other is to create an effective potential due to quantized
Landau levels of the magnetic field. The bound state energies of the
electron in L-shaped or T-shaped quantum wires are found to depend
quadratically (linearly) on the magnetic field in the weak (strong) field
region and are independent of the direction of the magnetic field. A simple
model is proposed to explain the behavior of the magnetic dependence of the
bound state energy both in weak and strong magnetic field regions.
\end{abstract}

\pacs{71,23,An;71.24.+q}
\maketitle

\address{Department of Electrophysics, National Chiao Tung\\
University 1001 Ta Hsueh Road, Hsinchu, 30050 Taiwan}

\address{Department of Electrophysics, National Chiao Tung University,
Hsinchu 300, Taiwan}

%\draft
%\wideabs{

% version ZI

%}

%%%%%%%%%%%%%%%%%%%%%%%%%%%%%%%%%%%%%%%%%%%%%%%%%%%%%%%%%%

\section{Introduction}

Recently, quasi-1D structures, such as quantum wires attract much attention
due to the enhanced confinement of the reduced dimension and the possibility
of tailoring the electronic and optical properties in applications\cite%
{Walck97}--\cite{Grundmann98}. Among the structures considered, the opened
quantum dot (OQD) is one of the simpler mesoscopic systems in which the
essential physics can be studied in great details. An OQD can be formed by
additional lateral confinements\cite{Liang97,Liang98} or by applying certain
magnetic fields\cite{Solimany95,Sim98}. Electrons and holes are trapped at
the L-shaped or T-shaped intersections because the single-particle
confinement energy can be found to be lower in the intersection of the arms.
These OQDs are quite different from the traditional quantum dots, since
there remain openings in such OQDs. Electrons in OQD systems are classically
unbounded. However, recent experimental photoluminescence spectroscopy
analyses\cite{Walck97,Glutsch97,Langbein96} have manifested that there are
bound states in such OQDs. The existence of bound states in OQDs essentially
shows the confinement effect of the mesoscopic geometry in quantum
mechanical region.

The exploration of the properties of bound states is a key to understand
some recent optical and electrical experiments on T-shaped quantum wires and
quantum dots\cite%
{Glutsch97,Langbein96,Brinkmann97,Grundmann98,Liang97,Liang98}. The
magneto-photoluminescence of T-shaped wires were measured recently\cite%
{Someya95}. The energy shift $\Delta E$ of PL peaks with magnetic field $B$
applied perpendicular to the wire axis and parallel to the stem wire was
measured. In these experiments, the information of exciton binding energy
can be provided from the photoluminescence spectroscopy. However, it is
unable to identify exactly the exciton binding energies unless we have the
knowledge of the confinement energy of either an electron or a hole in
quantum wires or quantum dots. Because they can not be extracted directly
from magneto-optical data due to the nonlinearity of the systems. In a
theoretical calculation of magneto-excitons in T-shaped wires\cite{Bryant01}%
, the observed field dependence of the exciton states for weak confinement
was reproduced, however, the diamagnetic shifts calculated from perturbation
theory is fail to describe the experimental results.

In this work, we consider two-dimensional OQDs which are formed at the
intersection of the arms of L-shaped or T-shaped quantum wires when
additional magnetic fields are applied perpendicular to the plane of arms. A
T-shaped quantum wire can be obtained by first growing a $%
GaAs/Al_xGa_{1-x}As $ superlattice on a (001) substrate, after cleavage, a $%
GaAs$ quantum wire is grown over the exposed (110) surface, resulting in a
T-shaped region where the electron or hole can be confined on a scale of
5-10 nm. The bound state energy of a charged particle (e.g. electron) in
such an opened quantum dot will be affected by the asymmetric geometry of
the system and the applied inhomogeneous magnetic fields. Intuitively, when
the confinment along one arm of the quantum wire is increased, confinment
along the orthogonal arm will decrease, because squeezing the electron or
hole in one arm will result in pushing the electron or hole out of the
quantum wire through the other arm. These pheonomena are not only
interesting in physics but also have no classical correspondence. To our
knowledge this squeezing effect have not been studied thoroughly.
Furthermore, T-shaped semiconductor quantum wires could be exploted as
three-terminal quantum interference devices, thus the study on the L-shaped
or T-shaped quantum wire is also important in practical applications.

\section{Formulation}

In the present work, a two-dimensional T-shaped (TOQW) or L-shaped opened
quantum wire (LOQW) is considered. A quantum dot with an area of $%
W_{1}\times W_{2}$ is formed in the intersection region while magnetic
fields $B_{1}$, $B_{2}$ and $B_{3}$ are applied perpendicularly to the other
subregions of the TOQW as shown in Fig. 1(a). The LOQW as shown in Fig. 1(b)
can be regarded as a transformation of TOQW in which the arm 2 is cut off. 
\begin{figure}[h]
\includegraphics[width=8cm]{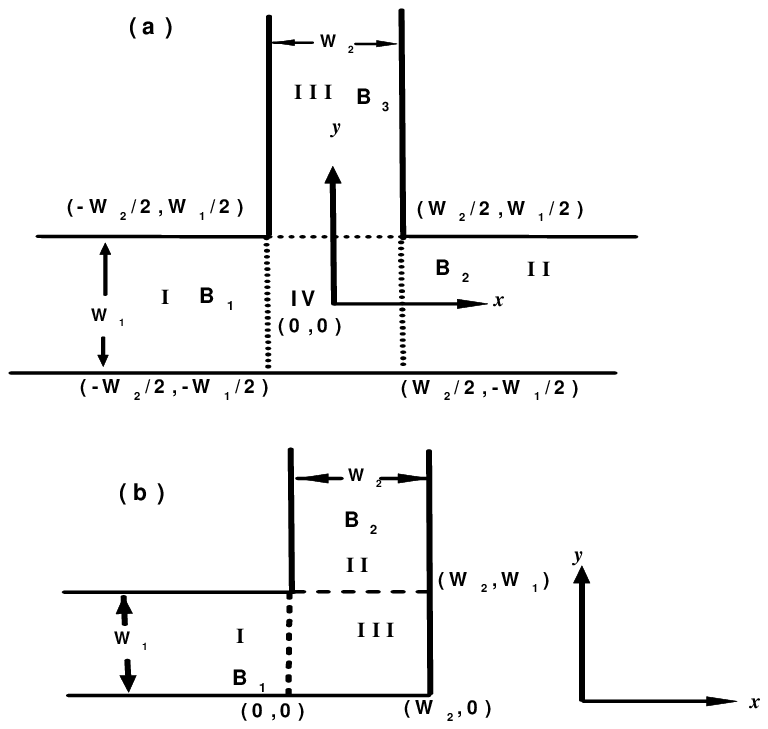}
\caption{The illustrations of the geometries of OQDs in (a) TOQW and (b)
LOQW systems.}
\end{figure}
For simplicity, the boundaries are assumed to be a hard--wall confinement
potential, leading to the formation of a magnetically confined cavity in
which the confinement of electron is enhanced. The transverse potential
inside the TOQW or LOQW is assumed to be zero. The magnetic fields are
assumed to be uniform in each individual subregion. Landau gauge is chosen
for the vector potential in different subregions: 
\begin{eqnarray*}
\mathbf{{A}(x,y)} &\mathbf{=}& \\
&&\mathbf{\left\{ 
\begin{array}{ll}
\begin{array}{l}
(0,B_{1}(x+0.5W_{2}))=(-B_{1}y,0) \\ 
+\nabla B_{1}(x+0.5W_{2})y,%
\end{array}
& \mbox{in region I;} \\ 
\begin{array}{l}
(0,B_{2}(x-0.5W_{2})=(-B_{2}y,0) \\ 
+\nabla B_{2}(x-0.5W_{2})y,%
\end{array}
& \mbox{in region II;} \\ 
\begin{array}{l}
(-B_{3}(y-0.5W_{1}),0)=(0,B_{3}x) \\ 
-\nabla B_{3}x(y-0.5W_{1}),%
\end{array}
& \mbox{in region III}; \\ 
(0,0), & \mbox{in
region IV.}%
\end{array}%
\right. }
\end{eqnarray*}%
The form of gauge guarantees the continuity of the vector potential at each
interface. The origin is chosen at the center of the intersection region.The
wavefunctions of the bound state $n$ of an electron for different subregions 
$I,II,III.IV$ are 
\begin{equation*}
\begin{array}{ll}
\begin{array}{l}
\label{Beq:psi1}\Psi _{n}^{I}=e^{-i(x+0.5W_{2})yeB_{1}/\hbar } \\ 
\times \left[ \sum_{m}r_{mn}e^{ik_{m}^{I}(x+0.5W_{2})}\Phi _{m}^{I}(y)\right]%
\end{array}
& \mbox{in region I;} \\ 
\begin{array}{l}
\Psi _{n}^{II}=e^{-i(x-0.5W_{2})yeB_{2}/\hbar } \\ 
\times \left[ \sum_{m}t_{m}e^{ik_{mn}^{I}(x-0.5W_{2})}\Phi _{m}^{II}(y)%
\right]%
\end{array}
& \mbox{in region II;} \\ 
\begin{array}{l}
\Psi _{n}^{III}=e^{ix(y-0.5W_{1})eB_{3}/\hbar } \\ 
\times \left[ \sum_{m}s_{mn}e^{ik_{m}^{II}(y-0.5W_{1})}\Phi _{m}^{III}(x)%
\right]%
\end{array}
& \mbox{in region III;}%
\end{array}%
\end{equation*}%
\begin{eqnarray}
\Psi _{n}^{IV} &=&\sum_{j}f_{j}(y)\left[ a_{jn}\sin k_{j}^{\prime
}(x-0.5W_{2})+b_{jn}\sin k_{j}^{\prime }(x+0.5W_{2})\right]  \notag
\label{Beq:psi4} \\
&&+c_{jn}g_{j}(x)\sin k_{j}^{^{\prime \prime }}(y+0.5W2)
\end{eqnarray}%
where 
\begin{eqnarray}
f_{j}(y) &=&\sqrt{\frac{2}{W_{1}}}\sin (\frac{j\pi }{W_{1}}y),-0.5W_{1}\leq
y\leq 0.5W_{1} \\
g_{j}(x) &=&\sqrt{\frac{2}{W_{2}}}\sin (\frac{j\pi }{W_{2}}x).-0.5W_{2}\leq
x\leq 0.5W_{2}
\end{eqnarray}%
$k_{j}^{\prime }=[k^{2}-(j\pi /W_{1})]^{1/2}\,$,$\,k_{j}^{^{\prime \prime
}}=[k^{2}-(j\pi /W_{2})]^{1/2}$ and $k_{m}^{i},\,i=I,II,III,\cdots $. Now
drop the subscript $n$ and substitute Eqs.(2) into the Schr\"{o}dinger
equation. After solving it numerically, one obtains eigen-wave-numbers $%
\{k_{m}^{I}\}$, $\{k_{m}^{II}\}$, $\{k_{m}^{III}\}$, the expansion
coefficients in Eqs. (\ref{Beq:psi1}) and (\ref{Beq:psi4}), and the
eigen-wave-functions $\{\Phi _{m}^{I}(y)\}$, $\{\Phi _{m}^{II}(x)\}$, $%
\{\Phi _{m}^{III}(x)\}$.\newline

\section{Results and Discussions}

\begin{figure}[h]
\includegraphics[width=8cm]{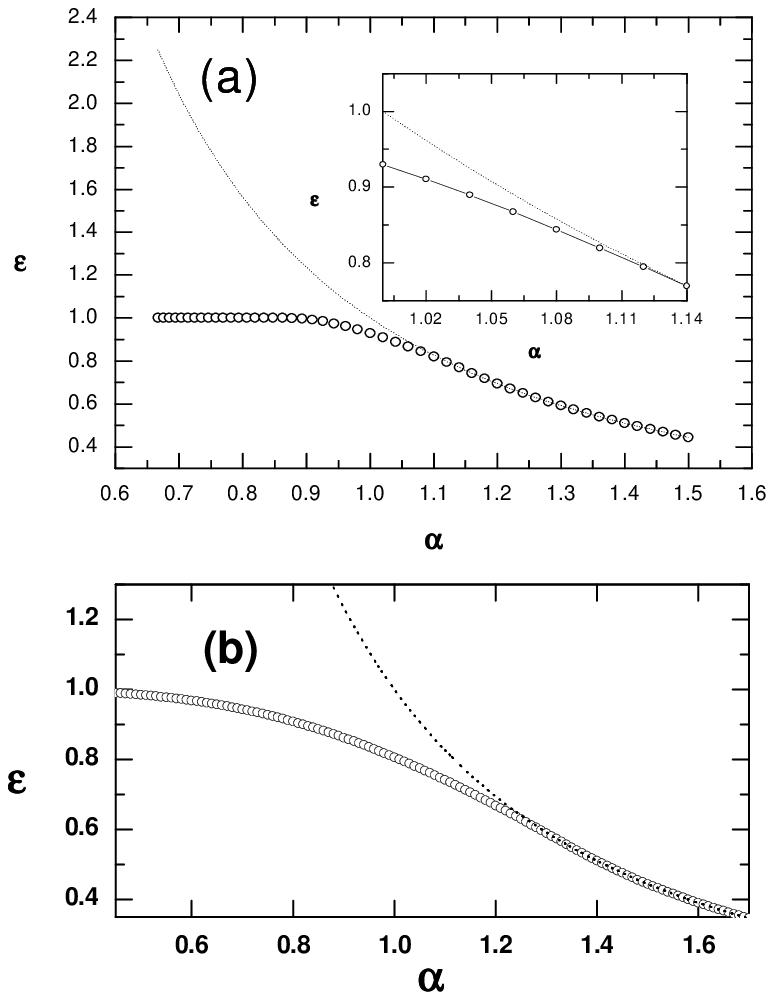}
\caption{(a).The bound state energy $\protect\varepsilon $ versus the
asymmetric ratio $\protect\alpha =W_{2}/W_{1}$ at zero magnetic field
strength. Open circle is our result. The dotted line is the curve $1/\protect%
\alpha $ as a guide to eyes. $E_{1}=\frac{\hbar ^{2}\protect\pi ^{2}}{%
2m^{\ast }W_{1}^{2}}$ is the first threshold energy of arm 1 (the region I).
(b).The bound state energy $\protect\varepsilon $ of a TOQW plotted in unit
of $E_{1}$ as a function of $\protect\alpha $. The bound state energy of the
electron approaches to unity for $\protect\alpha <<1$ and can be
approximately expressed by the curve $1/\protect\alpha ^{2}$ for $\protect%
\alpha \geq 1.33.$}
\end{figure}
Fig. 2(a) presents the calculated bound state energy of an electron in a
LOQW as a function of arm ratio $\alpha $. The bound state energy of the
electron is expressed in terms of the dimensionless quantity $\varepsilon
=E/E_{1}$, where $E_{1}=\frac{\hbar ^{2}\pi ^{2}}{2m^{\ast }W_{1}^{2}}$ is
the first subband energy in arm 1. One can note from the figure that the
bound state energy becomes smaller as the arm ratio $\alpha $ becomes
larger. For $\alpha =1$ (i.e. $W_{1}=W_{2}$), $r_{m}=t_{m}$ at zero magnetic
field, the bound state energy is $0.92964E_{1}$. The bound state energy $%
\varepsilon $ goes down and behaves like the curve $1/\alpha ^{2}$ as the $%
\alpha $ is increased larger than $1.14$. A deviation from the curve $%
1/\alpha ^{2}$ is observed in the region of $\alpha \leq 1.14$ as shown in
the inset of Fig. 2(a). The result can be ascribed to the fact that the
bound state energy of the electron matches the subband energy of arm 2 due
to the lateral confinement of region II. Since in this circumstance, $%
1/\alpha ^{2}(\pi /W_{1})^{2}$ is equal to $(\pi /W_{2})^{2}$, which is the
first subband level of the vertical wire. As the width $W_{2}$ becomes
larger and larger, the energy level becomes lower and lower, and gradually
coincides with the bound state level of the electron. Thus electron is
unable to be bounded in the corner region any more. As the asymmetry becomes
more prominently, the electronic energy becomes larger than the bottom of
the subband of the wider arm. However, if the energy of the electron state
is less than or just equal to the subband bottom, the electron is still
bounded inside the corner and does not move to the right or to the left. 
\begin{figure}[h]
\includegraphics[width=8cm]{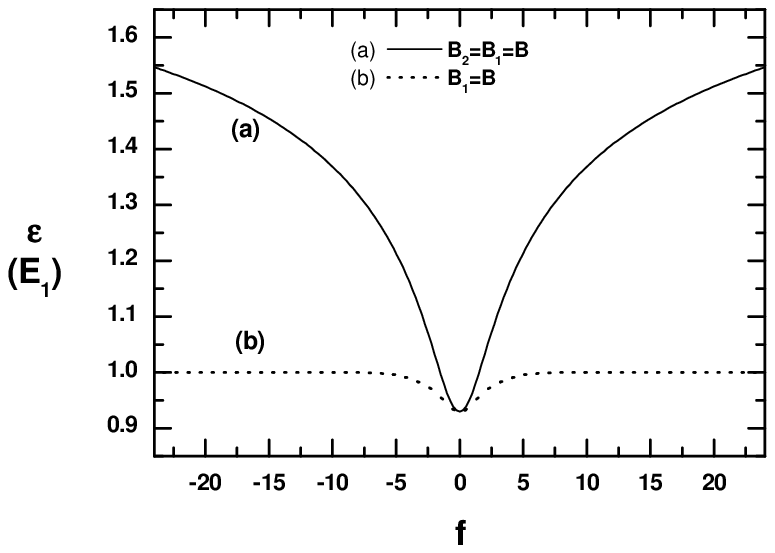}
\caption{ The bound state energy $\protect\varepsilon $ versus the field
strength $f$. (a) for both arms being acted by the magnetic fields in LOQW
system. (b) for only one arm being acted by the magnetic fields. The
dimensionless field strength $f$ is normalized by $E_{1}$.}
\end{figure}

Fig. 2(b) shows the bound state energy of the electron in a TOQW as a
function of $\alpha $. The bound state energy approaches unity as the width
of the vertical arm becomes very small, and behaves like the curve $1/\alpha
^{2}$ while $\alpha $ becomes larger. This is similar to the case of a LOQW.
The reason of this result can be understood intuitively that the
wavefunction of the electron is purged out of the vertical arm when it
becomes very narrow, therefore, the energy of this state is close to the
first threshold energy $E_{1}$ of the horizontal arm with a width of $W_{1}$%
. This bound state of the electron exists as long as the vertical arm is
infinite long, and is expected to disappear owing to the effect of leakage
if the arms is finite in length.

The calculated bound state energy of a symmetric LOQW in magnetic fields as
a function of the field strength $f=\hbar \omega _{c}/E_{1}$ is shown in
Fig. 3(a) and (b), where $\omega _{c}$ is cyclotron frequency of the
electron. One can observe that the bound state always exists when the
magnetic field is applied to both arms. The bound state depends linearly on
the magnetic field in weak field region while quadratically in strong field
region. increases monotonically as the magnetic field increases. However,
the energy of the bound state is pushed up by the applied magnetic field,
and then it goes up to $E_{1}$ when the magnetic field is applied to only
one arm. Thus, the electron can escape via the field free arm. Fig. 4
presents the confinement energy in a symmetric TOQW versus the field
strength when (a) all arms are acted by the same magnetic field $B$, (b) the
two horizontal arms are acted by the same magnetic field, and (c) only the
vertical arm is acted by the magnetic field. The same quadratic dependence
of magnetic field of the bound state energy is revealed again for weak
field, and the linear dependence appears in the strong field region as the
case of LOQW. Obviously, the bound state of the electron in a TOQW system
locates deeper than that in a LOQW, thus, the TOQW system has a weaker
confinement potential than the LOQW system.

The magnetic fields introduce a depleting effect on electrons and add an
extra potential surrounding the intersection region. The effective
potentials introduced by the magnetic fields are $k$--dependent. For the
bound state, these effective potentials are complex due to the pure
imaginary \{$k$\}. One expects intuitively that the magnetic field adds the
lowest Landau level $\frac{\hbar \omega _{c}}{2}=\frac{\hbar eB}{2m^{\ast }}$
directly to the quantum dot system an extra potential. Such levels are added
into the wire arm regions. However, the field plays another role due to the
essential physics of the magnetism. Qualitatively, one can understand the
effect induced by the magnetic field on the bound state by considering an
one-dimensional shallow quantum well with finite height $U_{0}$. In the
limit of shallow well, there is only one bound state exists in the well. Its
level energy is given by $E_{0}=U_{0}-(m^{\ast }W^{2}/2\hbar ^{2})U_{0}^{2}$%
, which is near the top of the well. As the magnetic field applies to the
system, the bound state energy changes because the potential height is
changed to $U_{0}+\frac{1}{2}\hbar \omega _{c}$. The variation of the state
level depends linearly on the potential height, i.e. 
\begin{equation*}
\frac{\partial E_{0}}{\partial U_{0}}=1-\frac{m^{\ast }W^{2}}{\hbar ^{2}}%
U_{0},
\end{equation*}%
The variation of the state level by taking account the depletion effect of
the magnetic field is assumed as 
\begin{equation*}
\frac{\partial E_{0}}{\partial W}=-\frac{m^{\ast }W}{\hbar ^{2}}U_{0}^{2}.
\end{equation*}%
Obviously, once we take the well shrunk into account, the quadratic form of
the dependence of magnetic field has to be considered also. This simple
model manifests the important geometric effect and the essential properties
of magnetism at the same time. Since the shrinking of the geometric scale is
no longer prominent in strong magnetic field region, the influence of the
magnetic field on the electron becomes smaller. Thus, the bound state energy
depends simply on the added effective potential, such that it seems likely
to depend linearly on the magnetic field in the strong magnetic field
region. 
\begin{figure}[h]
\includegraphics[width=8cm]{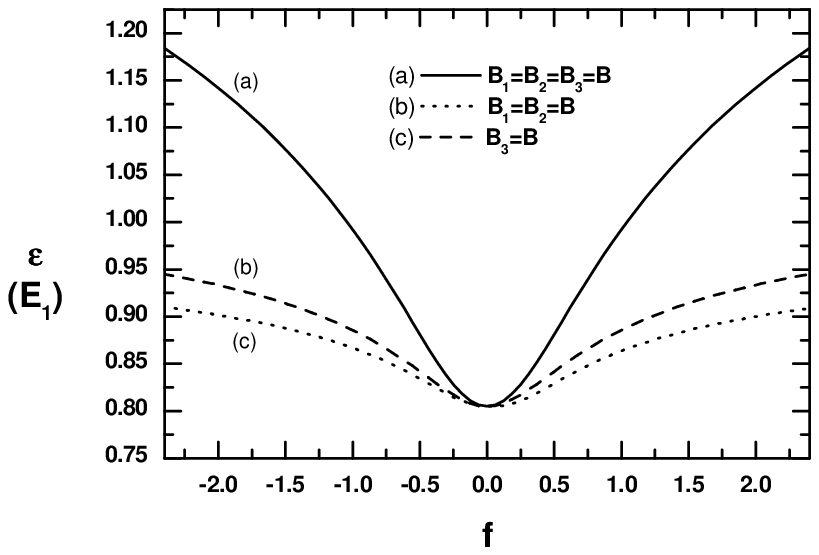}
\caption{ The bound state energy $\protect\varepsilon $ of T-shaped QW as a
function of the field strength $f$. Curve (a) for all arms being acted by
magnetic fields. The field strength $f$ is normalized by $E_{1}$. Curve (b)
for the horizontal arms being acted by magnetic fields, and curve (c) for
only vertical arm being acted by magnetic field. The dimensionless field
strength $f$ is normalized by $E_{1}$. }
\end{figure}

\section{Summary}

The effects of the asymmetric geometry and surrounding inhomogeneous
magnetic fields on the bound state of L-shaped or T-shaped quantum wires are
studied. When $\alpha $ increases, the bound state energy of the electron is
lower as expected. On the other hand, when the applied magnetic field
increases, the bound state level of the electron is pushed higher and higher
and the electron begins to be unbounded if there is an arm with finite
length which offers a passway for electron to leak out. Generally, the bound
state level of an electron in a TOQW system is lower than that in LOQW
system. This fact reflects the weaker confinement of the geometry. Parabolic
dependence of the bound state energy of the electron in weak field region on
the field strength is understood as a result of the depletion effect. In the
contrast, linear dependence in high field region is found to be resulted
from the additional effective potential due to the magnetic field.\newline

This work is supported partially by National Science Council, Taiwan under
the grant number NSC90-2112-M-009-018.

\end{document}